\documentclass{article}
\usepackage[T1]{fontenc} 
\usepackage[utf8]{inputenc} 
\usepackage{ismir,amsmath,cite,url}
\usepackage{graphicx}
\usepackage{color}
\usepackage{multirow}
\usepackage{booktabs}
\usepackage{amssymb}

\usepackage{balance}

\title{Exploring Post-Hoc Disentanglement of \\ Audio Foundation Models for Music Similarity}

\threeauthors
 {Lucas de Souza Brandão} {Universidade de São Paulo \\ {\tt lucasouza@usp.br}}
 {Gabriel da Costa Merlin} {Universidade de São Paulo \\ {\tt gabrielcmerlin@usp.br}}
 {Diego Furtado Silva} {Universidade de São Paulo \\ {\tt diegofsilva@usp.br}}

\def\authorname{L. Brandão, G. Merlin, and D. Silva}

\usepackage[bookmarks=false,pdfauthor={\authorname},pdfsubject={\papersubject},hidelinks]{hyperref}

\begin{document}

\maketitle

\begin{abstract}

Audio foundation models like CLAP, MuQ, and MERT excel at capturing general musical similarity. On the other hand, their entangled latent spaces limit interpretability and fine-grained attribute control. Conditional Similarity Networks (CSN) address this by pairing a backbone representation model with learned masks that gate the embedding space into attribute-specific subspaces. However, existing methods train the backbone from scratch, rather than leveraging pretrained foundation models. We instead present a systematic study of post-hoc disentanglement, adapting CSN to gate frozen pretrained embeddings. We evaluate adding trainable parameters to adapt the embedding under different masking and pooling regimes across two datasets. Our findings show that minimal adaptation capacity is sufficient for post-hoc gating to improve both attribute-specific disentanglement and overall retrieval performance, while pooling, masking, and depth choices yield limited additional gains.

\end{abstract}

\section{Introduction}

Musical similarity is inherently multi-faceted, as two tracks can be perceived as similar because they share characteristics such as genre, instrumentation, or emotional tone. Capturing this multiplicity requires representations rich enough to encode all these facets at once, making recent audio foundation models like CLAP~\cite{laion-clap}, MuQ~\cite{zhu2025muq}, and MERT~\cite{mert} natural candidates. Learned from large-scale pretraining, these foundation models excel at capturing overall musical similarity. Indeed, they capture perceptually dominant features like melody, closely aligning with human similarity judgments~\cite{papaioannou2026cross}. However, because these models collapse distinct musical characteristics into a single undifferentiated latent space, they lack disentanglement. In other words, the embedding cannot isolate or independently manipulate individual similarity factors. This makes it difficult to perform fine-grained tasks, such as retrieving tracks that match on one specific attribute, such as instrumentation, while differing on others. It also hinders the ability to explain why two tracks were retrieved as similar in the first place~\cite{afchar2022explainability}. Disentangling these factors is essential not only for model interpretability, but also for building controllable music retrieval systems~\cite{lee2020disentangled}.

Conditional Similarity Networks (CSN)~\cite{CSN} offer a disentanglement mechanism by learning masks that gate the embedding space provided by a backbone model according to a chosen musical factor. CSN and related conditional or attribute-specific embedding approaches have already been applied to music similarity, exploring different learning objectives and architectures for disentangled similarity. Specifically, existing methods condition subspaces on factors like genre, mood, instrumentation \cite{lee2020disentangled}, key, tempo, era~\cite{ribecky2021multi}, and structure representations across different hierarchy levels~\cite{buisson2022learning}. Furthermore, recent approaches extend these disentangled similarity spaces to specific instrumental parts from mixed signals~\cite{hashizume2025learning}. These works, however, train their backbones directly from scratch on audio features. This approach discards the knowledge already captured by large pretrained foundation models and requires computational resources that are not always accessible. This raises the question of whether disentanglement can instead be accomplished post hoc by adapting or masking an embedding space from a pretrained model, rather than training a disentangled representation from scratch.

In this work, we empirically study post-hoc disentangled representation learning for musical similarity built on pretrained audio foundation models (CLAP, MuQ, and MERT). We begin with CSN, adapting it to gate the pretrained embedding space into attribute-conditioned subspaces for genre, instrument, and mood without touching the backbone, and systematically explore different lightweight strategies to adapt and mask the embedding space. We evaluate disentanglement across two datasets to assess generalization beyond the training distribution. We investigate the impact of adding trainable parameters to adapt the embedding under different masking and pooling regimes, aiming to obtain representations that are both disentangled and effective for similarity retrieval without requiring heavy retraining.
\section{Proposed Method}
\label{sec:method}

We adapt CSN variants to extract disentangled, attribute-conditioned representations for instrument, genre, and mood features from frozen audio backbones. Temporal pooling first aggregates features into global vectors. An adaptation module then transforms these vectors before applying CSN-like masks, which route them into distinct attribute-specific subspaces. We optimize all parameters end-to-end using a regularized triplet loss.

\subsection{Backbones and Temporal Pooling}

We extract frame-level representations from frozen pretrained audio backbones (CLAP, MuQ, or MERT), yielding a temporal sequence of shape $[T, D_{\text{emb}}]$, where $T$ denotes the number of temporal frames and $D_{\text{emb}}$ is the backbone embedding dimension. Since $T$ remains large for multi-minute tracks, making downstream processing memory-intensive, we downsample the temporal axis via adaptive average pooling into 128 uniform segments. This yields a fixed sequence shape of $[128, D_{\text{emb}}]$, standardizing representations across tracks of varying durations.

\subsection{Adaptation Modules and Masking Strategies}

To aggregate the temporal sequence $[128, D_{\text{emb}}]$ into a single $D_{\text{emb}}$-dimensional vector, we evaluate two architectural options prior to attribute gating: (i) non-parametric \textbf{Mean Pooling}, which averages the 128 temporal tokens, or (ii) parametric \textbf{Attention Pooling}, which learns attention scores over the 128 temporal tokens via a linear layer, normalizes them across time via Softmax, and computes a weighted sum to emphasize salient temporal regions.

The aggregated vector $\mathbf{x}$ is passed through one of three adaptation modules: \textbf{Identity}, which leaves $\mathbf{x}$ unchanged; the \textbf{One Layer} module, which applies LayerNorm, a linear projection, GELU activation, and a residual connection; and the \textbf{MLP} module, which applies LayerNorm, two linear projections with an intermediate GELU, dropout, and a residual connection. We denote the resulting vector by $\hat{\mathbf{x}}$.

To capture distinct attributes, we define a condition-specific mask vector $\mathbf{m}_c \in \mathbb{R}^{D_{emb}}$ for each condition $c \in \mathcal{C} = \{\text{genre}, \text{instrument}, \text{mood}\}$. We obtain attribute-conditioned embeddings through element-wise gating:

\begin{equation}
\mathbf{e}_c = \mathbf{m}_c \odot \hat{\mathbf{x}}.
\end{equation}

We follow \cite{CSN} and compare two masking regimes. \textbf{Fixed Masks} use deterministic, non-overlapping binary masks $\mathbf{m}_c \in \{0,1\}^{D_{emb}}$ that define equally sized, contiguous partitions of embedding dimensions and remain frozen during training. \textbf{Learned Masks} are initialized randomly as $\mathbf{m}_c \sim \mathcal{N}(0.9, 0.7^2)$, updated via gradient descent, and passed through ReLU before gating. 

\subsection{Objective Function}

The framework is optimized by combining a primary metric learning objective with auxiliary regularization terms that encourage mask sparsity, subspace orthogonality, and embedding stability. For a given condition $c \in \mathcal{C}$, the training signal is enforced via a margin-based triplet loss in the masked subspace:

\begin{equation}
\mathcal{L}_{\text{triplet}}^{(c)} = \max\left(0,\ d(\mathbf{e}_c^a, \mathbf{e}_c^p) - d(\mathbf{e}_c^a, \mathbf{e}_c^n) + \Delta\right),
\end{equation}

where $d(\cdot, \cdot)$ denotes Euclidean distance, $\Delta$ is the margin, and $\mathbf{e}_c^a, \mathbf{e}_c^p, \mathbf{e}_c^n$ represent the masked embeddings of the anchor, positive, and negative tracks, respectively. This objective pulls the anchor closer to tracks sharing condition $c$ while pushing away non-matching tracks. 

To prevent mask values from growing unbounded and inflating the gated embeddings, we apply an $L_1$ penalty to each condition mask $\mathbf{m}_c$. This encourages sparse masks that retain only the dimensions relevant to each attribute:

\begin{equation}
\mathcal{L}_{\text{mask}}^{(c)} = \|\mathbf{m}_c\|_1.
\end{equation}

To reduce redundancy across different attribute subspaces and encourage disentanglement, we apply an orthogonality penalty across all condition mask pairs:

\begin{equation}
\mathcal{L}_{\text{ortho}}
=
\sum_{c_i \neq c_j \in \mathcal{C}}
\left(
\frac{\mathbf{m}_{c_i}^{\top}\mathbf{m}_{c_j}}
{\|\mathbf{m}_{c_i}\|_2\|\mathbf{m}_{c_j}\|_2}
\right)^2.
\end{equation}

To regularize the latent space and discourage arbitrarily large embedding norms that could artificially shrink the effective triplet distance, we apply an $L_2$ penalty to the unmasked representation $\hat{\mathbf{x}}$ before gating:

\begin{equation}
\mathcal{L}_{\text{emb}} = \|\hat{\mathbf{x}}\|_2^2.
\end{equation}

Finally, all remaining models are optimized end-to-end, using the combined training objective:

\begin{equation}
\mathcal{L} = \sum_{c \in \mathcal{C}} \left( \mathcal{L}_{\text{triplet}}^{(c)} + \lambda_{\text{mask}} \mathcal{L}_{\text{mask}}^{(c)} \right) + \lambda_{\text{ortho}} \mathcal{L}_{\text{ortho}} + \lambda_{\text{emb}} \mathcal{L}_{\text{emb}},
\end{equation}

where $\lambda_{\text{mask}}$, $\lambda_{\text{ortho}}$, and $\lambda_{\text{emb}}$ control the relative contribution of the auxiliary regularization terms. Under the Fixed Masks regime, $\mathcal{L}_{\text{mask}}^{(c)}$ and $\mathcal{L}_{\text{ortho}}$ remain constant and produce no gradients, leaving the adaptation module parameters as the only trainable parameters. Under the Identity module, $\mathcal{L}_{\text{emb}}$ also produces no gradients.
\section{Experimental Design}

To address the research questions, we designed the experiments described below, evaluating every combination of backbone, temporal pooling method, adaptation module, and masking regime introduced in Section~\ref{sec:method}.

\begin{table*}[ht]
\centering
\footnotesize
\resizebox{17.2cm}{!}{
    \begin{tabular}{l cc cc cc}
      \toprule
      & \multicolumn{2}{c}{\textbf{CLAP}} & \multicolumn{2}{c}{\textbf{MERT}} & \multicolumn{2}{c}{\textbf{MuQ}} \\
      \cmidrule(lr){2-3} \cmidrule(lr){4-5} \cmidrule(lr){6-7}
      \textbf{Configuration} &
      \textbf{Jamendo} & \textbf{MTT} &
      \textbf{Jamendo} & \textbf{MTT} &
      \textbf{Jamendo} & \textbf{MTT} \\
      \midrule
      Baseline
        & 69.3 $\pm$ 0.0 & 74.8 $\pm$ 0.0
        & 58.1 $\pm$ 0.0 & 60.3 $\pm$ 0.0
        & 64.3 $\pm$ 0.0 & 70.0 $\pm$ 0.0 \\
      Identity + Mean + Learned Mask
        & 71.6 $\pm$ 0.3 & 76.5 $\pm$ 0.1
        & 57.8 $\pm$ 0.2 & 60.2 $\pm$ 0.3
        & 64.6 $\pm$ 1.1 & 69.9 $\pm$ 0.8 \\
      Identity + Attention + Learned Mask
        & 70.0 $\pm$ 0.3 & 74.2 $\pm$ 0.4
        & 58.1 $\pm$ 0.3 & 60.5 $\pm$ 0.3
        & 64.0 $\pm$ 0.4 & 69.9 $\pm$ 0.6 \\
      One Layer + Mean + Fixed Mask
        & 71.5 $\pm$ 0.3 & 74.1 $\pm$ 0.4
        & 68.4 $\pm$ 0.7 & 70.0 $\pm$ 2.1
        & 71.0 $\pm$ 0.3 & 70.8 $\pm$ 0.6 \\
      One Layer + Mean + Learned Mask
        & 71.2 $\pm$ 0.4 & 74.2 $\pm$ 0.2
        & 68.4 $\pm$ 0.7 & 70.1 $\pm$ 1.7
        & 70.9 $\pm$ 0.3 & 71.5 $\pm$ 0.6 \\
      One Layer + Attention + Fixed Mask
        & 71.4 $\pm$ 0.2 & 74.0 $\pm$ 0.4
        & 67.6 $\pm$ 0.4 & 68.5 $\pm$ 1.8
        & 71.0 $\pm$ 0.3 & 71.2 $\pm$ 0.5 \\
      One Layer + Attention + Learned Mask
        & 71.4 $\pm$ 0.1 & 74.3 $\pm$ 0.2
        & 68.4 $\pm$ 0.4 & 68.4 $\pm$ 1.7
        & \textbf{71.2} $\pm$ 0.2 & \textbf{72.0} $\pm$ 0.5 \\
      MLP + Mean + Fixed Mask
        & \textbf{72.7} $\pm$ 0.2 & \textbf{76.9} $\pm$ 0.4
        & \textbf{68.6} $\pm$ 0.4 & \textbf{70.3} $\pm$ 0.9
        & 70.3 $\pm$ 0.2 & 69.1 $\pm$ 0.9 \\
      MLP + Mean + Learned Mask
        & 72.4 $\pm$ 0.2 & 76.4 $\pm$ 0.4
        & \textbf{68.6} $\pm$ 0.4 & 70.1 $\pm$ 0.4
        & 70.2 $\pm$ 0.1 & 70.2 $\pm$ 0.8 \\
      MLP + Attention + Fixed Mask
        & 72.7 $\pm$ 0.1 & 76.7 $\pm$ 0.4
        & 68.4 $\pm$ 0.3 & 69.8 $\pm$ 0.5
        & 70.2 $\pm$ 0.3 & 69.8 $\pm$ 0.6 \\
      MLP + Attention + Learned Mask
        & 71.2 $\pm$ 0.9 & 73.3 $\pm$ 1.9
        & 68.4 $\pm$ 0.4 & 69.4 $\pm$ 0.7
        & 70.2 $\pm$ 0.5 & 70.5 $\pm$ 0.7 \\
      \bottomrule
    \end{tabular}
}
\caption{Macro-average performance per triplet type and run of all configurations across the Jamendo and MTT datasets. Values are reported as mean $\pm$ run standard deviation, in percentages. Bold marks the best mean values.}
\label{tab:macro_avg_results}
\end{table*}

\subsection{Backbone Processing}

Because these backbones have maximum input sequence lengths (e.g., CLAP accepts at most 10 seconds of audio per pass), long tracks are processed by extracting features from non-overlapping 10-second chunks and concatenating them along the time axis. For a given track, MERT and MuQ yield temporal feature sequences directly in the shape $[T, D_{\text{emb}}]$, while CLAP outputs dual-stream window representations of shape $[2, T, D_{\text{emb}}]$, which we average along the first dimension to obtain a uniform sequence.

\subsection{Datasets and Data Curation}

We conduct evaluations across two standard music information retrieval datasets. The first, \textbf{MTG-Jamendo}~\cite{bogdanov2019mtg}, is used for both training and evaluation purposes. We utilize the \texttt{raw\_30s} preprocessing, in which each clip has a duration of at least 30 seconds, containing 55,701 tracks in total. Data is partitioned into training, evaluation, and test sets. Split partitions are strictly grouped by artist identity across over 3,500 artists to prevent data leakage.

The second dataset, \textbf{MagnaTagATune (MTT)}\footnote{We excluded corrupted tracks: IDs 57881, 55753, and 35644.}\cite{law2009evaluation}, comprises 25,863 clips with noisy raw tag annotations. We perform text normalization to unify morphological variants\footnote{Details on variant aggregation are available on our \href{https://github.com/sb-lucas/Post-Hoc-DRL}{\underline{GitHub}.}} (e.g., merging ``clasical'', ``classical'', and ``classic'') and manually group the top tags into \textit{genre}, \textit{instrument}, and \textit{mood} categories. As MTT provides no standard training or validation split for our setup, we use it only as a held-out test set to assess generalization beyond MTG-Jamendo.

\subsection{Triplet Mining and Optimization}

Training and evaluating the masked embeddings described above requires constructing triplets $(x^a, x^p, x^n)$ for each similarity condition $c \in \mathcal{C}$. Across both training and evaluation, all triplets satisfy a unified set of criteria: the anchor $x^a$ and positive $x^p$ share at least one tag in $c$, whereas negatives $x^n$ must possess at least one tag in $c$, share zero tags in $c$ with $x^a$, and originate from an artist distinct from both $x^a$ and $x^p$. For deterministic and reproducible testing, these triplets are generated offline per attribute $c$. During training, they are instead mined dynamically using a semi-hard strategy~\cite{ribecky2021multi} to adapt to the evolving embedding space.

\subsection{Optimization and Implementation Details}

All models are trained end-to-end using the Adam optimizer with a learning rate of $10^{-4}$, a batch size of 128, and cosine annealing scheduling. Training proceeds until convergence, monitored via early stopping on validation loss with a patience of 20 epochs and an improvement threshold of $\delta = 0$. The loss function hyper-parameters are set in all experiments to $\lambda_{\text{mask}} = 10^{-5}$, $\lambda_{\text{ortho}} = 10^{-3}$, and $\lambda_{\text{emb}} = 10^{-5}$, with a triplet margin of $\Delta = 0.1$.

To systematically evaluate the framework, we test a complete grid combining every backbone (CLAP, MuQ, MERT), temporal pooling method (Mean, Attention), adaptation module (Identity, One Layer, MLP), and masking regime (Fixed, Learned). Each configuration is executed across five independent seeds (0--4). For Identity configuration, only Learned Masks apply.

\subsection{Evaluation Protocol and Baselines}

To evaluate retrieval performance, we use \textbf{Triplet Accuracy} separately for each similarity condition, yielding three triplet types $c'$. We denote the resulting accuracy by $\mathcal{A}(e,c')$, where $e$ is an embedding vector. Given a fixed combination $\theta$ of backbone model, pooling method, adaptation module, and masking regime, we define the \textbf{Effect} $\mathcal{E}$ of a mask $m_c$ on triplets of type $c'$ as the change in triplet accuracy relative to the unmasked representation $\hat{\mathbf{x}}$:

\begin{equation}
\mathcal{E}(\theta, m_c, c') = \mathcal{A}(e_c, c') - \mathcal{A}(\hat{\mathbf{x}}, c').
\end{equation}

A well-disentangled mask should have a positive effect on its target condition ($c'=c$) and a negative effect on non-matching conditions. Therefore, we define the \textbf{Specificity} $\mathcal{S}$ score to measure whether a mask $m_c$ selectively improves retrieval for its target condition $c$ while degrading retrieval for non-matching conditions in $\mathcal{C}$:

\begin{equation}
\mathcal{S}(\theta, m_c, \mathcal{C}) = \mathcal{E}(\theta, m_c, c) - \sum_{c'' \in \mathcal{C} \setminus \{c\}} \mathcal{E}(\theta, m_c, c'').
\end{equation}

\begin{table*}[!htp]
\centering
\footnotesize
\resizebox{17.2cm}{!}{
    \begin{tabular}{l cc cc cc}
      \toprule
      & \multicolumn{2}{c}{\textbf{CLAP}} & \multicolumn{2}{c}{\textbf{MERT}} & \multicolumn{2}{c}{\textbf{MuQ}} \\
      \cmidrule(lr){2-3} \cmidrule(lr){4-5} \cmidrule(lr){6-7}
      \textbf{Configuration} &
      \textbf{Jamendo} & \textbf{MTT} &
      \textbf{Jamendo} & \textbf{MTT} &
      \textbf{Jamendo} & \textbf{MTT} \\
      \midrule
      Identity + Mean + Learned Mask
        & 2.14 $\pm$ 0.30 & 2.71 $\pm$ 0.42
        & -0.03 $\pm$ 0.15 & 0.07 $\pm$ 0.37
        & 0.49 $\pm$ 0.81 & 0.37 $\pm$ 1.07 \\
      Identity + Attention + Learned Mask
        & 0.35 $\pm$ 0.31 & 0.24 $\pm$ 0.36
        & -0.03 $\pm$ 0.21 & 0.24 $\pm$ 0.47
        & -0.13 $\pm$ 0.34 & 0.57 $\pm$ 0.84 \\
      One Layer + Mean + Fixed Mask
        & 7.37 $\pm$ 0.31 & 4.84 $\pm$ 0.40
        & 4.83 $\pm$ 0.70 & 5.74 $\pm$ 0.76
        & \textbf{7.69} $\pm$ 0.30 & 3.51 $\pm$ 0.78 \\
      One Layer + Mean + Learned Mask
        & 7.02 $\pm$ 0.32 & 4.60 $\pm$ 0.20
        & \textbf{5.07} $\pm$ 0.45 & 5.71 $\pm$ 0.41
        & 7.12 $\pm$ 0.41 & 3.81 $\pm$ 0.40 \\
      One Layer + Attention + Fixed Mask
        & \textbf{7.41} $\pm$ 0.21 & 5.30 $\pm$ 0.49
        & 4.13 $\pm$ 1.14 & 4.07 $\pm$ 0.29
        & 6.97 $\pm$ 0.33 & 3.59 $\pm$ 0.43 \\
      One Layer + Attention + Learned Mask
        & 7.12 $\pm$ 0.34 & 4.82 $\pm$ 0.24
        & 5.04 $\pm$ 0.65 & 4.71 $\pm$ 0.40
        & 6.77 $\pm$ 0.22 & 3.84 $\pm$ 0.32 \\
      MLP + Mean + Fixed Mask
        & 4.88 $\pm$ 0.41 & \textbf{6.03} $\pm$ 0.33
        & 3.31 $\pm$ 1.35 & 4.21 $\pm$ 2.01
        & 7.11 $\pm$ 0.05 & 4.66 $\pm$ 0.50 \\
      MLP + Mean + Learned Mask
        & 3.46 $\pm$ 0.53 & 4.27 $\pm$ 0.48
        & 4.55 $\pm$ 0.19 & \textbf{6.25} $\pm$ 0.41
        & 6.56 $\pm$ 0.31 & 5.27 $\pm$ 0.54 \\
      MLP + Attention + Fixed Mask
        & 4.70 $\pm$ 0.27 & 5.77 $\pm$ 0.43
        & 2.40 $\pm$ 0.31 & 2.84 $\pm$ 0.71
        & 7.08 $\pm$ 0.18 & \textbf{5.93} $\pm$ 0.59 \\
      MLP + Attention + Learned Mask
        & 6.29 $\pm$ 1.24 & 4.59 $\pm$ 0.40
        & 4.54 $\pm$ 0.23 & 5.83 $\pm$ 0.26
        & 6.64 $\pm$ 0.25 & 4.93 $\pm$ 0.42 \\
      \bottomrule
    \end{tabular}
}
\caption{Average Specificity performance of all masks and runs of the evaluated configurations across the Jamendo and MTT datasets. Values are reported as mean $\pm$ run standard deviation, in percentages. Bold marks the best mean values.}
\label{tab:specificity_results}
\end{table*}

\section{Results}

Table~\ref{tab:macro_avg_results} reports macro-averaged triplet accuracy across the three triplet types and five runs. We also report the baseline performance, given by the unmasked, mean-pooled embedding of each backbone. Performance increases most for MERT and MuQ, suggesting that these models are less effective at direct similarity retrieval than CLAP, which shows smaller improvements. The Identity module, which applies no adaptation, generally yields little to no improvement and can even underperform the baseline, while the One Layer and MLP modules yield substantially larger gains. Thus, post-hoc disentanglement training can improve general similarity retrieval performance. However, neither the choice of adaptation module, pooling method, nor masking regime shows a consistent effect. These findings highlight the importance of adding trainable adaptation parameters and indicate that masking alone is insufficient to reshape the representation for improved retrieval.

\begin{table}[ht]
\centering
\footnotesize
\setlength{\tabcolsep}{4pt}
\renewcommand{\arraystretch}{1.1}
\begin{tabular}{ll ccc ccc}
\toprule
\textbf{Model} & \textbf{Mask} &
\multicolumn{3}{c}{\textbf{Jamendo}} &
\multicolumn{3}{c}{\textbf{MTT}} \\
\cmidrule(lr){3-5} \cmidrule(lr){6-8}
&&
\textbf{Inst.} & \textbf{Genre} & \textbf{Mood} &
\textbf{Inst.} & \textbf{Genre} & \textbf{Mood} \\
\midrule

\multirow{4}{*}{\textbf{CLAP}}
& None
& 73.2 & 75.5 & 66.4
& 73.7 & 86.2 & 64.8 \\
& Inst.
& \textbf{74.1} $\uparrow$ & 71.3 $\downarrow$ & 63.7 $\downarrow$
& \textbf{73.9} $\uparrow$ & 81.0 $\downarrow$ & 58.2 $\downarrow$ \\
& Genre
& 71.6 $\downarrow$ & \textbf{76.8} $\uparrow$ & 65.2 $\downarrow$
& 71.3 $\downarrow$ & \textbf{86.5} $\uparrow$ & 63.5 $\downarrow$ \\
& Mood
& 65.6 $\downarrow$ & 69.6 $\downarrow$ & \textbf{67.2} $\uparrow$
& 70.1 $\downarrow$ & 81.3 $\downarrow$ & \textbf{70.3} $\uparrow$ \\

\midrule

\multirow{4}{*}{\textbf{MERT}}
& None
& 69.2 & 71.7 & \textbf{64.2}
& \textbf{65.8} & 79.6 & 63.6 \\
& Inst.
& \textbf{69.8} $\uparrow$ & 68.0 $\downarrow$ & 61.5 $\downarrow$
& 64.9 $\downarrow$ & 75.7 $\downarrow$ & 59.2 $\downarrow$ \\
& Genre
& 69.0 $\downarrow$ & \textbf{72.2} $\uparrow$ & 63.3 $\downarrow$
& 64.8 $\downarrow$ & \textbf{80.0} $\uparrow$ & 61.5 $\downarrow$ \\
& Mood
& 63.4 $\downarrow$ & 66.6 $\downarrow$ & 64.0 $\downarrow$
& 62.4 $\downarrow$ & 73.0 $\downarrow$ & \textbf{66.0} $\uparrow$ \\

\midrule

\multirow{4}{*}{\textbf{MuQ}}
& None
& 68.6 & 72.9 & \textbf{66.8}
& \textbf{65.4} & 78.0 & \textbf{64.5} \\
& Inst.
& \textbf{70.0} $\uparrow$ & 65.3 $\downarrow$ & 60.4 $\downarrow$
& \textbf{65.4} $\mathord{=}$ & 71.7 $\downarrow$ & 60.1 $\downarrow$ \\
& Genre
& 67.2 $\downarrow$ & \textbf{75.6} $\uparrow$ & 64.5 $\downarrow$
& 62.9 $\downarrow$ & \textbf{78.4} $\uparrow$ & 60.6 $\downarrow$ \\
& Mood
& 58.7 $\downarrow$ & 63.1 $\downarrow$ & 65.3 $\downarrow$
& 61.3 $\downarrow$ & 69.8 $\downarrow$ & 63.4 $\downarrow$ \\

\bottomrule
\end{tabular}
\caption{Performance of the MLP configuration with Mean Pooling and Fixed Masks across the Jamendo and MTT datasets. Values are reported as percentages for instrument, genre, and mood retrieval. Bold values indicate the highest performance for each metric and backbone. Arrows indicate increases ($\uparrow$) or decreases ($\downarrow$) relative to the corresponding None-mask configuration.}\label{tab:mask_results}
\end{table}

To evaluate disentanglement, Table~\ref{tab:specificity_results} reports the average Specificity score for each evaluated configuration, calculated by first averaging across the three masks and then across the five independent runs. Again, trained adaptation modules perform best, whereas further architectural design choices show inconclusive results and depend heavily on both the chosen backbone and evaluation dataset. This indicates that even a minimal number of trainable parameters is sufficient, yet necessary, to achieve both post-hoc disentanglement and retrieval performance gains.

We further examine triplet accuracy for each mask and triplet type in Table~\ref{tab:mask_results}, considering only the MLP adaptation module with mean pooling and fixed masks. Triplet accuracy by type shows that genre is consistently the easiest to retrieve, while mood is the hardest across backbones and datasets. Analysis by mask shows that performance is very similar across masks for every triplet type. Since the fixed masks retain only one third of the embedding through a disjoint partition, these results suggest that the pretrained latent space is highly entangled, with attribute-relevant information distributed redundantly across dimensions rather than confined to specific ones. Moreover, the largest performance effects come from decreases on non-matching triplet types rather than gains on the matching one, suggesting that masking primarily suppresses interfering information rather than injecting new signal. This supports our central hypothesis that post-hoc disentanglement of pretrained backbones is achievable by reorganizing existing representations without backbone retraining.

\section{Conclusion}

In this work, we present a post-hoc strategy to disentangle representations from frozen audio foundation models (CLAP, MuQ, and MERT) for attribute-conditioned music similarity retrieval using Conditional Similarity Networks. We evaluate the addition of trainable parameters to adapt the embeddings and test different pooling methods and masking regimes across two datasets. We find that the original embeddings are highly entangled and redundant, allowing high triplet accuracy across all conditions using any third of the representation. We show that introducing even minimal adaptation capacity improves both triplet accuracy and disentanglement, whereas direct mask training without additional parameters performs poorly. Other architectural variations yield inconclusive results. Specifically, these disentanglement gains arise primarily from suppressing cross-attribute interference rather than from improving performance on the corresponding similarity condition. Overall, our findings show that post-hoc disentanglement training is a promising strategy for effectively adapting frozen foundation models to attribute-based music retrieval without backbone retraining.

\section{Acknowledgments}

This study was financed, in part, by the São Paulo Research Foundation (FAPESP), Brasil. Process Number 2025/07753-1 and Process Number 2024/13907-9.

\bibliography{references}

@misc{CSN,
      title={Conditional Similarity Networks}, 
      author={Andreas Veit and Serge Belongie and Theofanis Karaletsos},
      year={2017},
      eprint={1603.07810},
      archivePrefix={arXiv},
      primaryClass={cs.CV},
      url={https://arxiv.org/abs/1603.07810}, 
}

@article{ribecky2021multi,
  title={Multi-input Architecture and Disentangled Representation Learning for Multi-dimensional Modeling of Music Similarity},
  author={Ribecky, Sebastian and Abe{\ss}er, Jakob and Lukashevich, Hanna},
  journal={arXiv preprint arXiv:2111.01710},
  year={2021}
}

@misc{mert,
      title={MERT: Acoustic Music Understanding Model with Large-Scale Self-supervised Training}, 
      author={Yizhi Li and Ruibin Yuan and Ge Zhang and Yinghao Ma and Xingran Chen and Hanzhi Yin and Chenghua Lin and Anton Ragni and Emmanouil Benetos and Norbert Gyenge and Roger Dannenberg and Ruibo Liu and Wenhu Chen and Gus Xia and Yemin Shi and Wenhao Huang and Yike Guo and Jie Fu},
      year={2023},
      eprint={2306.00107},
      archivePrefix={arXiv},
      primaryClass={cs.SD}
}

@misc{laion-clap,
      title={Large-scale Contrastive Language-Audio Pretraining with Feature Fusion and Keyword-to-Caption Augmentation}, 
      author={Yusong Wu and Ke Chen and Tianyu Zhang and Yuchen Hui and Marianna Nezhurina and Taylor Berg-Kirkpatrick and Shlomo Dubnov},
      year={2024},
      eprint={2211.06687},
      archivePrefix={arXiv},
      primaryClass={cs.SD},
      url={https://arxiv.org/abs/2211.06687}, 
}

@conference {mtg-jamendo,
    author = "Bogdanov, Dmitry and Won, Minz and Tovstogan, Philip and Porter, Alastair and Serra, Xavier",
    title = "The MTG-Jamendo Dataset for Automatic Music Tagging",
    booktitle = "Machine Learning for Music Discovery Workshop, International Conference on Machine Learning (ICML 2019)",
    year = "2019",
    address = "Long Beach, CA, United States",
    url = "http://hdl.handle.net/10230/42015"
}

@inproceedings{lee2020disentangled,
  title={Disentangled multidimensional metric learning for music similarity},
  author={Lee, Jongpil and Bryan, Nicholas J and Salamon, Justin and Jin, Zeyu and Nam, Juhan},
  booktitle={ICASSP 2020-2020 IEEE International Conference on Acoustics, Speech and Signal Processing (ICASSP)},
  pages={6--10},
  year={2020},
  organization={IEEE}
}

@article{hashizume2025learning,
  title={Learning separated representations for instrument-based music similarity},
  author={Hashizume, Yuka and Li, Li and Miyashita, Atsushi and Toda, Tomoki},
  journal={arXiv preprint arXiv:2503.17281},
  year={2025}
}

@inproceedings{buisson2022learning,
  title={Learning multi-level representations for hierarchical music structure analysis},
  author={Buisson, Morgan and Mcfee, Brian and Essid, Slim and Crayencour, Helene-Camille},
  booktitle={International Society for Music Information Retrieval (ISMIR)},
  year={2022}
}

@article{papaioannou2026cross,
  title={Cross-Cultural Music Similarity: Bridging Human Perception, Signal Processing, and Foundation Models},
  author={Papaioannou, Charilaos and Benetos, Emmanouil and Potamianos, Alexandros},
  journal={Transactions of the International Society for Music Information Retrieval},
  volume={9},
  number={1},
  pages={347--368},
  year={2026},
  publisher={Ubiquity Press}
}

@article{zhu2025muq,
  title={Muq: Self-supervised music representation learning with mel residual vector quantization},
  author={Zhu, Haina and Zhou, Yizhi and Chen, Hangting and Yu, Jianwei and Ma, Ziyang and Gu, Rongzhi and Luo, Yi and Tan, Wei and Chen, Xie},
  journal={IEEE Transactions on Audio, Speech and Language Processing},
  year={2025},
  publisher={IEEE}
}

@article{afchar2022explainability,
  title={Explainability in music recommender systems},
  author={Afchar, Darius and Melchiorre, Alessandro and Schedl, Markus and Hennequin, Romain and Epure, Elena and Moussallam, Manuel},
  journal={AI Magazine},
  volume={43},
  number={2},
  pages={190--208},
  year={2022}
}

@conference {bogdanov2019mtg,
    author = "Bogdanov, Dmitry and Won, Minz and Tovstogan, Philip and Porter, Alastair and Serra, Xavier",
    title = "The MTG-Jamendo Dataset for Automatic Music Tagging",
    booktitle = "Machine Learning for Music Discovery Workshop, International Conference on Machine Learning (ICML 2019)",
    year = "2019",
    address = "Long Beach, CA, United States",
    url = "http://hdl.handle.net/10230/42015"
}

@inproceedings{law2009evaluation,
  title={Evaluation of algorithms using games: The case of music tagging.},
  author={Law, Edith and West, Kris and Mandel, Michael I and Bay, Mert and Downie, J Stephen},
  booktitle={ISMIR},
  pages={387--392},
  year={2009}
}

\end{document}